\documentclass{emulateapj}

\renewcommand{\vec}[1]{\mathbf{#1}}
\newcommand{\eqref}[1]{(\ref{#1})}
\begin{document}
  
  \title{Numerical simulations of imbalanced strong
  magnetohydrodynamic turbulence} 
  \author{Jean Carlos Perez} 
  \email{jcperez@wisc.edu}
  \author{Stanislav Boldyrev}
  \email{boldyrev@wisc.edu}
  \affil{Department of Physics, University of Wisconsin at
  Madison, 1150 University Ave, Madison, WI 53706, USA}
  \date{\today}

\begin{abstract}
Magnetohydrodynamics (MHD) is invoked to address 
turbulent fluctuations in a variety of astrophysical systems. MHD
turbulence in nature is often anisotropic and imbalanced, in that
Alfv\'enic fluctuations moving in opposite directions along the
background magnetic field carry unequal energies.  This work formulates
specific requirements for effective numerical simulations of strong
imbalanced MHD turbulence with a guide field $\vec B_0$. High-resolution
simulations are then performed and they suggest that the spectra of the
counter-propagating Alfv\'en modes do not differ from the balanced case,
while their amplitudes and the corresponding rates of energy cascades
are significantly affected by the imbalance. It is further proposed that
the stronger the imbalance the larger the magnetic Reynolds number that
is required in numerical simulations in order to correctly reproduce the
turbulence spectrum. This may explain current discrepancies among
numerical simulations and observations of imbalanced MHD turbulence.
\end{abstract}

\maketitle

\section{Introduction}
Magnetohydrodynamic (MHD) turbulence plays a critical role in 
theoretical modeling of the spectra of velocity and magnetic
fluctuations in the Solar Wind
\citep[e.g.,][]{coleman66,coleman68,belcher71,marsch91,goldstein95} as
well as electron density fluctuations in the interstellar medium
\citep[e.g.,][]{armstrong95,lithwick01}. Following the pioneering works
of \cite{iroshnikov63} and \cite{kraichnan65}, and the introduction of
the concept of \emph{critical balance} by \cite{GS95}, progress in MHD
turbulence was mostly due to phenomenological models, closure theories, 
and weak turbulence asymptotic models. In recent years, constantly
increasing computing power 
has put within our
reach the ability to simulate universal inertial range spectra of strong MHD turbulence 
\citep[e.g.,][]{cho00,maron01,haugen03,biskamp,muller05,
mason06,mason08,mininni07,beresnyak08,perez08,perez09}.  These
simulations have motivated new phenomenological models and have revived
the vigorous interest in the fundamentals of MHD turbulence.

As an important result, simulations have shown that strong MHD turbulence is
locally \emph{imbalanced}, that is, it spontaneously develops correlated
regions of positive and negative cross-helicity where the energy in
Alfv\'en waves propagating along and against the magnetic field are not
equal, irrespective of the total amount of cross-helicity in the system,
see~\cite{perez09,mason09}. In each of these regions both energy and
cross helicity are subject to a nonlinear cascade from large to small
scales, consequently having an effect in the overall energy spectrum.

Such an imbalance is clearly present in the solar wind, where velocity
and magnetic fluctuations show high correlations of a preferred sign,
that is, the normalized \emph{cross helicity} $ \sigma_c =
{\left\langle\vec v\cdot\vec b\right\rangle}/{E}={H_c}/{E}$ is
predominantly close to unity. Here $E$ is the average total energy
(kinetic plus magnetic), $H_c$ is the \emph{cross helicity} and the
brackets denote a suitable ensemble average.  The preferred positive
sign of $\sigma_c$ indicates that there is more energy in Alfv\'en waves
propagating outwards from the Sun than propagating inwards.

Several phenomenological models have been proposed to address strong imbalanced
MHD
turbulence~(\citet{lithwick07,chandran08,beresnyak08,perez09,podesta09}),
some with support from numerical simulations and observations. However,
these works have led to conflicting predictions. For instance, the
theory by Lithwick et al. concludes that in imbalanced regions the
Els\"asser spectra have the same scalings $E^+(k_\perp)\propto
E^-(k_\perp)\propto k_\perp^{-5/3}$. The theory by Chandran proposes
that the spectra of $E^+(k_\perp)$ and $E^-(k_\perp)$ are different
depending of the degree of imbalance. The theory by Beresnyak \& Lazarian
also suggests the different spectra for $E^+(k_\perp)$ and
$E^-(k_\perp)$. Finally, the analysis by Perez \& Boldyrev and Podesta
\& Bhattacharjee finds that the spectra of $E^+(k_\perp)$ and
$E^-(k_\perp)$ have different amplitudes but the same scalings
$E^+(k_\perp)\propto E^-(k_\perp)\propto k_\perp^{-3/2}$.

On the numerical side, anisotropy and imbalance significantly complicate
simulations of MHD turbulence.  In the presence of a strong guide
field, the turbulent fluctuations become elongated in the direction of
the field. This increases the variety of numerical settings that can be
used to drive such turbulence and complicates comparison of numerical
simulations carried out by different groups. For example, numerical
settings can include weak or strong guide fields; different levels of
imbalance; cubic or rectangular simulation boxes; short or long-time
correlated large-scale forces; frozen large-scale modes; forcing that
excites only $\vec v$, $\vec b$ or both; volume forcing or forcing at
the boundaries of the domain; different resolutions along
field-perpendicular and field-parallel directions, etc.

To understand how different numerical simulations can be compared to
each other, one needs to understand the conditions that should be met to
ensure that simulations reproduce the range of scales where turbulence
is universal. For instance, one has to be sure that the forcing and the
dissipation do not spoil the inertial interval.  The first steps in this
direction were made in \citep{mason06,perez08} for the balanced case. It
was found that the inertial range scaling can be sensitive to the way in
which the turbulence is forced, not because of a lack of universality,
but because the inertial range has a limited extent.

In the present paper we formulate the conditions that need to be
satisfied in order to correctly simulate strong imbalanced MHD
turbulence.  We then propose a universal numerical setting, based on the
reduced MHD model, which allows one to find the spectra of both balanced
and imbalanced MHD turbulence.  We found that imbalance significantly
reduces the inertial interval in numerical simulations. The numerical
resolution required to produce a large inertial interval strongly
increases with the amount of imbalance. An imbalance essentially
stronger than $\gamma^2=(z^+)^2/(z^-)^2\sim 10$ cannot be accessed with
currently available supercomputers, which significantly limits the
applicability of present-day numerical simulations to practical cases.

\section{Model equations}
In the presence of a guide field ${\bf B_0}$ (say in the $z$ direction),
the MHD equations 
describing the evolution of magnetic and velocity fluctuations, ${\bf
b}(\vec x,t)$ and ${\bf v}(\vec x,t)$ can be written in terms of the
so-called Els\"asser variables, ${\bf z}^{\pm}={\bf v}\pm {\bf b}$:
\begin{eqnarray}
  \partial_t {\bf z}^{\pm} \mp ({\bf V}_A\cdot \nabla){\bf
    z}^{\pm}+({\bf z}^{\mp}\cdot \nabla){\bf z}^{\pm}=-\nabla P +{\bf
    f},
  \label{mhd1}
\end{eqnarray}
where ${\bf V}_{A}={\bf B}_{0}/\sqrt{4\pi \rho}$ is the Alfv\'en
velocity, $\rho$ is the fluid density, $P$ is the total pressure that is
determined from the incompressibility condition, $\nabla \cdot {\bf
z}^{\pm}=0$, ${\bf f}$ is large-scale forcing and we neglect small
viscosity and resistivity.  
In the limit of weak fluctuations, the incompressible MHD system
describes non-interacting Alfv\'en waves with dispersion relation
$\omega^\pm(\vec k)=\pm k_\| V_A$. These waves can have two types of
polarizations, the shear-Alfv\'en one ($\vec z^\pm_{S}$) and the
pseudo-Alfv\'en one ($\vec z^\pm_{P}$) given in Fourier space by $\vec
e_S\equiv {\vec e_\|\times\vec k}/{k_\perp}$ and $\vec e_P\equiv {\vec
k\times\vec e_S}/k $, where $\vec e_\|$ is a unit vector in $z$
direction.  Strong MHD turbulence is dominated by 
fluctuations with $k_\perp \gg k_\|$.  \citet{GS95} argued that since
for large $k_{\perp}$ the polarization of the pseudo-Alfv\'en
fluctuations is almost parallel to the guide field, such fluctuations
are coupled only to field-parallel gradients, which are small since
$k_\| \ll k_{\perp}$. Therefore, the pseudo-Alfv\'en modes do not play a
dynamically essential role in the turbulent cascade. We can remove the
pseudo-Alfv\'en modes by setting $\vec z^\pm_\|=0$ in
equations~\eqref{mhd1} to obtain
\begin{equation}   
  \displaystyle \partial_t {\tilde {\bf z}}^{\pm} \mp ({\bf V}_A\cdot
    \nabla){\tilde {\bf z}}^{\pm}+({\tilde {\bf z}}^{\mp}\cdot
    \nabla){\tilde {\bf z}}^{\pm}=-\nabla_{\perp} P +\frac
    1R\nabla^2{\tilde {\bf z}}^{\pm},
  \label{rmhd} 
\end{equation}
where $R$ denotes the Reynolds numbers (discussed below).  In this
system, the fluctuating fields have only two vector components, ${\tilde
{\bf z}}^{\pm}=\{{\tilde z}^{\pm}_1, {\tilde z}^{\pm}_2, 0 \}$, but
depend on all three spatial coordinates. 
System (\ref{rmhd}) is equivalent to the Reduced MHD model, originally
developed for tokamak plasmas by \cite{kadomtsev74} and
\cite{strauss76}.

\section{Numerical strategy}\label{numstrategy}
Model~(\ref{rmhd}) describing the nonlinear interactions of
Shear-Alfv\'en modes is a valuable tool in theoretical and numerical
studies of incompressible MHD turbulence. Based on this model, we now
discuss the conditions that should be satisfied in order to correctly
simulate strong MHD turbulence, both balanced and imbalanced.

\subsection{Computation Domain} 
In the presence of a strong guide field the turbulence is
anisotropic. According to \citet{GS95}, deep in the inertial range
turbulence should become strong and the fluctuations should satisfy the
critical balance condition $k_{\|}B_0\sim k_\perp b_{\lambda}$, where
$k_\|\sim 1/l$ and $k_\perp\sim 1/\lambda$ are the field-parallel and
field-perpendicular wave vectors associated to an anisotropic eddy of
parallel size $l$ and perpendicular size $\lambda$, respectively.  Let
us define the nonlinear interaction strength parameter
\begin{equation}
  \chi=(k_\perp b_{\lambda})/(k_\|B_0);
\end{equation}
the critical balance condition then implies~$\chi\sim 1$.

Simulations of incompressible turbulence are generally based on the
Fourier pseudo-spectral method. In order to allow for the inertial
interval to develop, turbulence is driven at the lowest resolvable wave
numbers, and the energy dissipates at large wave numbers determined by
the Reynolds numbers. For simulations on a cubic periodic box of size
$L$, the smallest wave-numbers along the field-parallel and
field-perpendicular directions coincide, i.e., $k_\perp =
k_\|=2\pi/L$. Therefore, driving at the low $k_\|,k_\perp$ results in an
isotropic forcing and the nonlinear strength parameter at the forcing
scale becomes $ \chi_0 = {k_\perp b_{\lambda}}/{k_\|B_0}\sim
{b_{\lambda}}/{B_0}\ll 1 $, which means that at least at the large
scales, nonlinear interactions are weak. This would not be harmful if we
had the resources to achieve arbitrarily high resolution, as the
turbulence would proceed weakly until $\chi\sim 1$, and then would
become strong. However, simulations generally produce a rather limited
inertial range, so that the parameter $\chi$ can hardly reach unity in
such a setup.

As pointed out by \cite{maron01}, and applied in recent
simulations~\cite{mason06,perez08}, the best way to avoid this is to use
an anisotropic domain such that at the forcing scale the parameter
$\chi$ is already of order unity, that is, the excited large-scale modes
are already anisotropic and satisfy the critical balance condition.  To
achieve this we choose an elongated box $L_\perp^2\times L_\|$, so that
the lowest field-perpendicular and field-parallel wave-numbers are
$k_\perp = 2\pi/L_\perp$ and $k_\|=2\pi/L_\|$, respectively. In this
case, forcing at the lowest $k_\perp,k_\|$ leads to $
\chi_0 = L_\| b_{\lambda}/L_\perp B_0, $ which is of order unity
provided that
\begin{equation}
{L_\perp}/{L_\|}\sim {b_{\lambda}}/{B_0}.
  \label{crit_bal}
\end{equation}
In this way, the turbulence is excited in a strong regime and the
cascade proceeds down to smaller scales preserving the critical balance
condition.

\subsection{Numerical Resolution}
At first sight, it appears that elongating the box along the $z$
direction to match the elongation of the eddies should not change the
number of grid points required in this 
direction compared to the number of points in the $x$ and $y$
directions. Fortunately, the number of points in the $z$ direction can be
reduced. This follows from the fact that the turbulent spectrum declines
quite slowly, as a power-law, in the $k_\perp$ direction, while it drops
sharply in the $k_\|$ direction for $k_\|> k_{\perp}^{\alpha}$, where
$\alpha$ is a some positive power not exceeding~1
\citep{cho00,maron01,oughton04,perez08,perez09} . This qualitatively different
spectral behavior in $k_\|$ and $k_\perp$ directions allows one to
reduce the numerical resolution by a factor of 2 to 4 in the parallel
direction, see Table~{\ref{sims_table}}. We checked that the restoration
of the full resolution in the $z$ direction does not change the results,
while significantly increases the computing costs.

\subsection{Periodic Boundary Conditions} The spectral method assumes
periodic boundary conditions in all spatial directions. The periodic
boundary conditions in the z-direction may cause questions of whether
the Alfv\'en modes counter-propagating along a given magnetic field line
will always interact only among themselves. The answer is no, since
periodicity of the fluctuations does not imply periodicity of the
magnetic field lines. Magnetic field lines are not periodic, and each
given eddy interacts with many independent counter-propagating
eddies. One can check numerically, that reducing the parallel box size
$L_\|$ below (\ref{crit_bal}) somewhat spoils the spectrum at low wave
numbers, e.g., as seen in forced simulations of \citep{muller05}, while
increasing it beyond (\ref{crit_bal}) does not change the results, but
increases the computational cost [Mason \& Cattaneo, unpublished].

\subsection{Reynolds numbers}\label{re} Probably the most significant limitation
is imposed by the consideration of imbalanced turbulence. Indeed, in the
imbalanced case, $\gamma =z^+/z^-> 1$, the formal Reynolds numbers
corresponding to $\vec z^+$ and $\vec z^-$ fields are essentially
different. Therefore, the resolution requirements increase with the
amount of imbalance in order to produce large inertial ranges. Assume
that the number of grid points in a field-perpendicular direction scales
with the Reynolds number as $N\sim Re^{\beta}$, where $\beta=2/3$ or
$3/4$ depending on the spectral slope ($3/2$ or $5/3$). Then increasing
the imbalance $\gamma$ by, say, a factor of 3 will require increasing
the resolution by approximately a factor of 2. Noting that resolution of
at least 1024 is required to simulate the imbalance $\gamma\sim 2$ (see
below), we conclude that significantly stronger imbalance is not
achievable with present day computing power.

\subsection{Random Forcing} 
In this section we discuss the important
aspects to be considered when choosing a particular forcing. We assume a
random force ${\tilde {\bf f}}$ that has no component along $z$, it is
solenoidal in the $x-y$ plane and its Fourier coefficients outside the
range
$ 1 \leq k_{\perp} \leq 2, \quad(2\pi/L_\|) \leq k_\| \leq
  (2\pi/L_\|)n_z\label{modes} $
are zero, where $n_z$ determines the width of the force spectrum in
$k_\|$, and $L_\perp = 2\pi$. The Fourier coefficients inside that range
are Gaussian random numbers with amplitude chosen so that the resulting
rms velocity fluctuations are of order unity.  The individual random
values are refreshed independently at time intervals~$\tau$.  The
parameter~$n_z$ controls the degree to which the critical balance
condition is satisfied at the forcing scale. Note that we do not drive
the $k_\|=0$~mode but allow it to be generated by nonlinear
interactions.

In contrast with incompressible hydrodynamic system, an incompressible
MHD system can support Alfv\'en waves. When the system is driven by a
time dependent forcing, the most effectively driven modes are those 
resonating with the frequencies present in the forcing. Therefore, the spatial spectra of the large-scale velocity and
magnetic fluctuations are not generally the same as the spatial spectrum of the force. 
Rather, they essentially depend on the {\em both} spatial {\em and} temporal spectra of the random forcing. 
Ultimately, it is the spectrum of the large-scale fluctuations, not the driving force, 
that should be controlled in numerical simulations.

In the following we perform simulations with a short-time-correlated
random forcing that drives the turbulence close to
the critical balance condition at the large scales. The short
correlation time $\tau$ ensures a broadband frequency spectrum for the
forcing, which allows high frequency Alfv\'en waves to be excited, so
that one can control the width of the field-parallel spatial spectrum of
the fluctuations by choosing the number of driven field-parallel
modes,~$n_z$. A sort-time correlated Gaussian random force has another
important advantage. It allows one to control the rates at which the
energies of $z^+$ and $z^-$ modes are injected.
\begin{table}[!tb]
\begin{tabular}{c@{\hspace{0.5cm}}c@{\hspace{0.5cm}}c@{\hspace{0.5cm}}c@{\hspace{0.5cm}}c@{\hspace{0.5cm}}c@{\hspace{0.5cm}}c@{\hspace{0.5cm}}} \hline\hline 
    Run & Resolution         & $Re$ & $\nu=\eta $ & $L_\|/L_\perp$ & $\sigma_c$ \\ 
    \hline 
    A1  & $ 512^2\times 256$ & 2400 &$4.2\times 10^{-4}$ & 5           & 0   \\  
    A2  & $1024^2\times 256$ & 6000 &$1.7\times 10^{-4}$ & 5           & 0   \\
    B1  & $ 256^3$           & 900  &$1.1\times 10^{-3}$ & 10          & 0.6 \\
    B2  & $ 512^2\times 256$ & 2200 &$4.6\times 10^{-4}$ &10               & 0.6 \\ 
    B3  & $1024^2\times 256$ & 5600 &$1.8\times 10^{-4}$ &10               & 0.6 \\ 
    \hline\hline
  \end{tabular}
  \caption{Summary of simulations of strong balanced turbulence (A1, A2) and strong imbalanced turbulence (B1, B2, B3).}\label{sims_table}
\end{table}

\section{Numerical results}
Table \ref{sims_table} summarizes five representative simulations that
incorporate all the aspects discussed in section \ref{numstrategy}. The
simulations produce consistent and physically meaningful results for a
range of Reynolds numbers and degrees of imbalance. Runs A are carried out 
for balanced turbulence, that is, $\sigma_c=0$. In runs B, cross
helicity is injected at the forcing scale in such a way that $\sigma_c$ 
reaches a steady state of $\sim 0.6$. We
use a short time-correlated forcing, compared to the Alfv\'en time of
the excited modes, so that the energy injection rates for both $\vec
z^+$ and $\vec z^-$ only depend on the variance of the imposed forcing,
which is controlled in our simulations. In the imbalanced
case, field-parallel box size is optimized to reach the critical balance at the
large scales. Except for the Reynolds numbers, simulations B1, B2, and B3
have the exact same parameters including the energy injection rates,
$\epsilon^+$ and $\epsilon^-$.

\begin{figure}[!th]
  \begin{center}
    \includegraphics[width=0.5\textwidth]{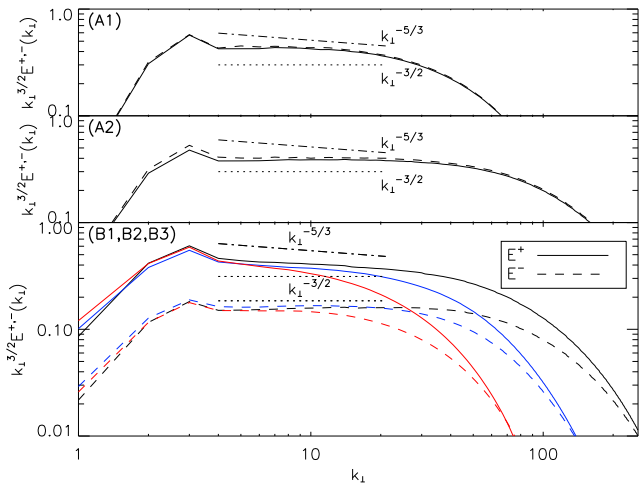}
  \end{center}
  \caption{Spectra of the Els\"asser fields in numerical simulations of MHD turbulence. Top two
    frames: balanced turbulence (runs A1, A2); bottom frame: imbalanced
    turbulence (runs B1-B3).}\label{strong}
\end{figure}

Since the background magnetic field must be strong, we choose $B_0=5$ in the
$v_{rms}$ units, cf the discussion in~\citep[][]{mason06}. Time is
normalized to the large scale eddy turnover time $\tau_0=L_\perp/2\pi
v_{rms}$, where $L_\perp$ is the field-perpendicular box size. The
Reynolds number is defined as $Re=v_{rms}(L_\perp/2\pi)/\nu$ and we have
chosen the same value for the magnetic Reynolds number,
$Rm={b}_{rms}(L_\perp/2\pi)/\eta$, denoting both by~$R$
in~(\ref{rmhd}). In each run, the average is performed over about 100
large-scale-eddy turnover times. The results are presented in Fig.~(\ref{strong}). 

\section{Discussion} 

The advantage of our optimized setup (reduced MHD, elongated box,
reduced resolution in z-direction) can be seen already in simulations 
of balanced turbulence, top frames in Fig.~(\ref{strong}), runs A1 and
A2. The energy spectra approach $E^\pm(k_\perp)\propto k_\perp^{-3/2}$
in good agreement with earlier numerical findings
\citep[e.g.,][]{maron01,haugen03,muller05,mason06}, while our
simulations require considerably less computational cost and produce
slightly larger 
inertial intervals. Our most significant results are obtained for
the imbalanced case.  The bottom frame of Fig.~\ref{strong} shows the
spectra for three different Reynolds numbers (runs B1 to B3). It is
observed that the spectra $E^\pm$ are pinned at the
dissipation scales, which supports the phenomenological predictions
by \cite{grappin83,galtier00,lithwick03,chandran08}. We also find that the
large-scale parts of the both spectra are practically insensitive to the
Reynolds numbers. These two important properties imply that as the Re
numbers are further increased, the $E^\pm$ spectra must become
progressively more parallel in the inertial interval. This is indeed
seen in our numerical simulations (B1-B3). Moreover, our numerical
simulations indicate that both spectra approach the universal scaling of
strong MHD turbulence $E^\pm(k_\perp)\propto k_\perp^{-3/2}$, while they
have essentially different amplitudes and correspond to essentially
different energy fluxes.


The proposed numerical setup and the results of our simulations can
serve as an important tool for resolving currently existing
controversies regarding the spectra of imbalanced MHD turbulence. In
particular, our results are consistent with the theories and
observations predicting same scaling $-3/2$ for both Els\"asser fields
\citep{perez09,podesta09}. They are also broadly consistent with the
theory by \citet{lithwick07}, predicting the same scaling $-5/3$ for both
fields. The major difference between the two phenomenologies is in the
phenomenon of dynamic alignment that is taken into account in the former
models and is not considered in the latter one. Our numerical findings
are less consistent with the models predicting different scalings for
the Els\"asser fields $E^{\pm}$,
\citep[e.g.,][]{chandran08,beresnyak08}. It would be interesting to
understand what assumptions of these models disagree with the
numerics. Finally, our analysis may explain somewhat puzzling
numerical findings by \citet[][]{beresnyak09}, who report different
spectra for the $E^{\pm}$ Els\"asser fields, and the intersection of the
spectra rather than pinning at the dissipation scale. According to our
results, the explanation might lie in the fact that in these
simulations the imbalance was extremely high, up to
$\gamma^2=(z^+)^2/(z^-)^2 \sim 1000$, and, therefore, the universal
regime of imbalanced MHD turbulence was not reached, see our analysis in
Sec.~\ref{re}.\\

This work was supported by the U.S. DOE grants No.~DE-FG02-07ER54932 and
DE-SC0001794, and by the NSF Center for Magnetic
Self-Organization in Laboratory and Astrophysical Plasmas at the
University of Wisconsin-Madison.  High Performance Computing 
resources were provided by the Texas Advanced Computing Center (TACC) at
the University of Texas at Austin under the NSF-Teragrid Project
TG-PHY080013N.


\begin{thebibliography}{32}
\expandafter\ifx\csname natexlab\endcsname\relax\def\natexlab#1{#1}\fi

\bibitem[{{Armstrong} {et~al.}(1995){Armstrong}, {Rickett}, \&
  {Spangler}}]{armstrong95}
{Armstrong}, J.~W., {Rickett}, B.~J., \& {Spangler}, S.~R. 1995, \apj, 443, 209

\bibitem[{{Belcher} \& {Davis}(1971)}]{belcher71}
{Belcher}, J.~W. \& {Davis}, Jr., L. 1971, \jgr, 76, 3534

\bibitem[{{Beresnyak} \& {Lazarian}(2008)}]{beresnyak08}
{Beresnyak}, A. \& {Lazarian}, A. 2008, \apj, 682, 1070

\bibitem[{{Beresnyak} \& {Lazarian}(2009)}]{beresnyak09}
---. 2009, \apj, 702, 460

\bibitem[{{Biskamp}(2003)}]{biskamp}
{Biskamp}, D. 2003, {Magnetohydrodynamic Turbulence}, ed. D.~{Biskamp}

\bibitem[{{Boldyrev} {et~al.}(2009){Boldyrev}, {Mason}, \&
  {Cattaneo}}]{mason09}
{Boldyrev}, S., {Mason}, J., \& {Cattaneo}, F. 2009, \apjl, 699, L39

\bibitem[{{Chandran}(2008)}]{chandran08}
{Chandran}, B.~D.~G. 2008, \apj, 685, 646

\bibitem[{{Cho} \& {Vishniac}(2000)}]{cho00}
{Cho}, J. \& {Vishniac}, E.~T. 2000, \apj, 539, 273

\bibitem[{{Coleman}(1966)}]{coleman66}
{Coleman}, P.~J. 1966, PRL, 17, 207

\bibitem[{{Coleman}(1968)}]{coleman68}
{Coleman}, Jr., P.~J. 1968, \apj, 153, 371

\bibitem[{{Galtier} {et~al.}(2000){Galtier}, {Nazarenko}, {Newell}, \&
  {Pouquet}}]{galtier00}
{Galtier}, S., {Nazarenko}, S.~V., {Newell}, A.~C., \& {Pouquet}, A. 2000,
  Journal of Plasma Physics, 63, 447

\bibitem[{{Goldreich} \& {Sridhar}(1995)}]{GS95}
{Goldreich}, P. \& {Sridhar}, S. 1995, \apj, 438, 763

\bibitem[{{Goldstein} {et~al.}(1995){Goldstein}, {Roberts}, \&
  {Matthaeus}}]{goldstein95}
{Goldstein}, M.~L., {Roberts}, D.~A., \& {Matthaeus}, W.~H. 1995, \araa, 33,
  283

\bibitem[{{Grappin} {et~al.}(1983){Grappin}, {Leorat}, \&
  {Pouquet}}]{grappin83}
{Grappin}, R., {Leorat}, J., \& {Pouquet}, A. 1983, \aap, 126, 51

\bibitem[{{Haugen} {et~al.}(2003){Haugen}, {Brandenburg}, \&
  {Dobler}}]{haugen03}
{Haugen}, N.~E.~L., {Brandenburg}, A., \& {Dobler}, W. 2003, \apjl, 597, L141

\bibitem[{{Iroshnikov}(1963)}]{iroshnikov63}
{Iroshnikov}, P.~S. 1963, \azh, 40, 742

\bibitem[{{Kadomtsev} \& {Pogutse}(1974)}]{kadomtsev74}
{Kadomtsev}, B.~B. \& {Pogutse}, O.~P. 1974, JETP, 38, 283

\bibitem[{{Kraichnan}(1965)}]{kraichnan65}
{Kraichnan}, R.~H. 1965, Physics of Fluids, 8, 1385

\bibitem[{{Lithwick} \& {Goldreich}(2001)}]{lithwick01}
{Lithwick}, Y. \& {Goldreich}, P. 2001, \apj, 562, 279

\bibitem[{{Lithwick} \& {Goldreich}(2003)}]{lithwick03}
---. 2003, \apj, 582, 1220

\bibitem[{{Lithwick} {et~al.}(2007){Lithwick}, {Goldreich}, \&
  {Sridhar}}]{lithwick07}
{Lithwick}, Y., {Goldreich}, P., \& {Sridhar}, S. 2007, \apj, 655, 269

\bibitem[{{Maron} \& {Goldreich}(2001)}]{maron01}
{Maron}, J. \& {Goldreich}, P. 2001, \apj, 554, 1175

\bibitem[{{Marsch}(1991)}]{marsch91}
{Marsch}, E. {MHD turbulence in the solar wind.}, 159--241

\bibitem[{{Mason} {et~al.}(2006){Mason}, {Cattaneo}, \& {Boldyrev}}]{mason06}
{Mason}, J., {Cattaneo}, F., \& {Boldyrev}, S. 2006, PRL, 97, 255002

\bibitem[{{Mason} {et~al.}(2008){Mason}, {Cattaneo}, \& {Boldyrev}}]{mason08}
---. 2008, \pre, 77, 036403

\bibitem[{{Mininni} \& {Pouquet}(2007)}]{mininni07}
{Mininni}, P.~D. \& {Pouquet}, A. 2007, PRL, 99, 254502

\bibitem[{{M{\"u}ller} \& {Grappin}(2005)}]{muller05}
{M{\"u}ller}, W.-C. \& {Grappin}, R. 2005, PRL, 95, 114502

\bibitem[{{Oughton} {et~al.}(2004){Oughton}, {Dmitruk}, \&
  {Matthaeus}}]{oughton04}
{Oughton}, S., {Dmitruk}, P., \& {Matthaeus}, W.~H. 2004, Physics of Plasmas,
  11, 2214

\bibitem[{{Perez} \& {Boldyrev}(2008)}]{perez08}
{Perez}, J.~C. \& {Boldyrev}, S. 2008, \apjl, 672, L61

\bibitem[{{Perez} \& {Boldyrev}(2009)}]{perez09}
---. 2009, PRL, 102, 025003

\bibitem[{{Podesta} \& {Bhattacharjee}(2009)}]{podesta09}
{Podesta}, J.~J. \& {Bhattacharjee}, A. 2009, ArXiv e-prints

\bibitem[{{Strauss}(1976)}]{strauss76}
{Strauss}, H.~R. 1976, Physics of Fluids, 19, 134

\end{thebibliography}

\end{document}